%% file: Manuscript_Spirale.tex
\documentclass{nature}
\usepackage{amsmath,amssymb}
\usepackage[dvipsnames]{xcolor } 
\usepackage{graphicx}
\usepackage{dcolumn}
\usepackage{bm}

\makeatletter
\let\saved@includegraphics\includegraphics
\AtBeginDocument{\let\includegraphics\saved@includegraphics}
\renewenvironment*{figure}{\@float{figure}}{\end@float}
\makeatother

\title{Giving light a new twist with standard optical fibres: \\rainbow Archimedean spiral emission}

\author{F. Mangini$^1*$, M. Ferraro$^2*$, M. Zitelli$^2*$, V. Kalashnikov$^2$, A. Niang$^1$, T. Mansuryan$^3$, F. Frezza$^2$, A. Tonello$^3$, V. Couderc$^3$, A.B. Aceves$^4$, S. Wabnitz$^{2}$}

\begin{document}

\maketitle

\begin{affiliations}
 \item Department of Information Engineering (DII), University of Brescia, Via Branze 38, 25123 Brescia, Italy.
 \item Department of Information Engineering, Electronics and Telecommunications (DIET),
Sapienza University of Rome, Via Eudossiana 18, 00184 Rome, Italy.
 \item Universit\'e de Limoges, XLIM, UMR CNRS 7252, 123 Avenue A. Thomas, 87060 Limoges, France
 \item Department of Mathematics, Southern Methodist Universit,3100 Dyer St, Dallas, TX 75205, USA\\
$^*$ These authors have contributed equally
\end{affiliations}

\begin{abstract}
We demonstrate a new practical approach for generating multicolour spiral-shaped beams. It makes use of a standard silica optical fibre, combined with a titled input laser beam. The resulting breaking of the fibre axial symmetry leads to the propagation of a helical beam. The associated output far-field has spiral shape, independently of the input laser power value. Whereas, with a high-power near-infrared femtosecond laser, a visible supercontinuum spiral emission is generated. With appropriate control of the input laser coupling conditions, the colours of the spiral spatially self-organize in a rainbow distribution. Our method is independent of the laser source wavelength and polarization. Therefore, standard optical fibres may be used for generating spiral beams in many applications, ranging from communications to optical tweezers and quantum optics.
\end{abstract}
\bigskip
\section*{Introduction}
Spiral shaped beams are conventionally obtained by means of spiral phase plates\cite{khonina2020properties} or by interfering Gaussian waves with the so-called longitudinal orbital angular momentum (LOAM) carrying beams\cite{zhang2017orbital}. 
LOAM beams are peculiar solutions of Maxwell's equations, characterized by phase singularity and helical wavefront 
 $\varphi$\cite{wright2015controllable,zhang2019generation}. The main characteristic of LOAM beams is the so-called topological charge ($\ell$), an integer number that counts the number of wave front rotations over one wavelength of propagation, so that the electromagnetic field has a phase $\phi=\ell \varphi$.

\begin{figure}[!ht]
	  \centering
	  \includegraphics[width=0.7\textwidth]{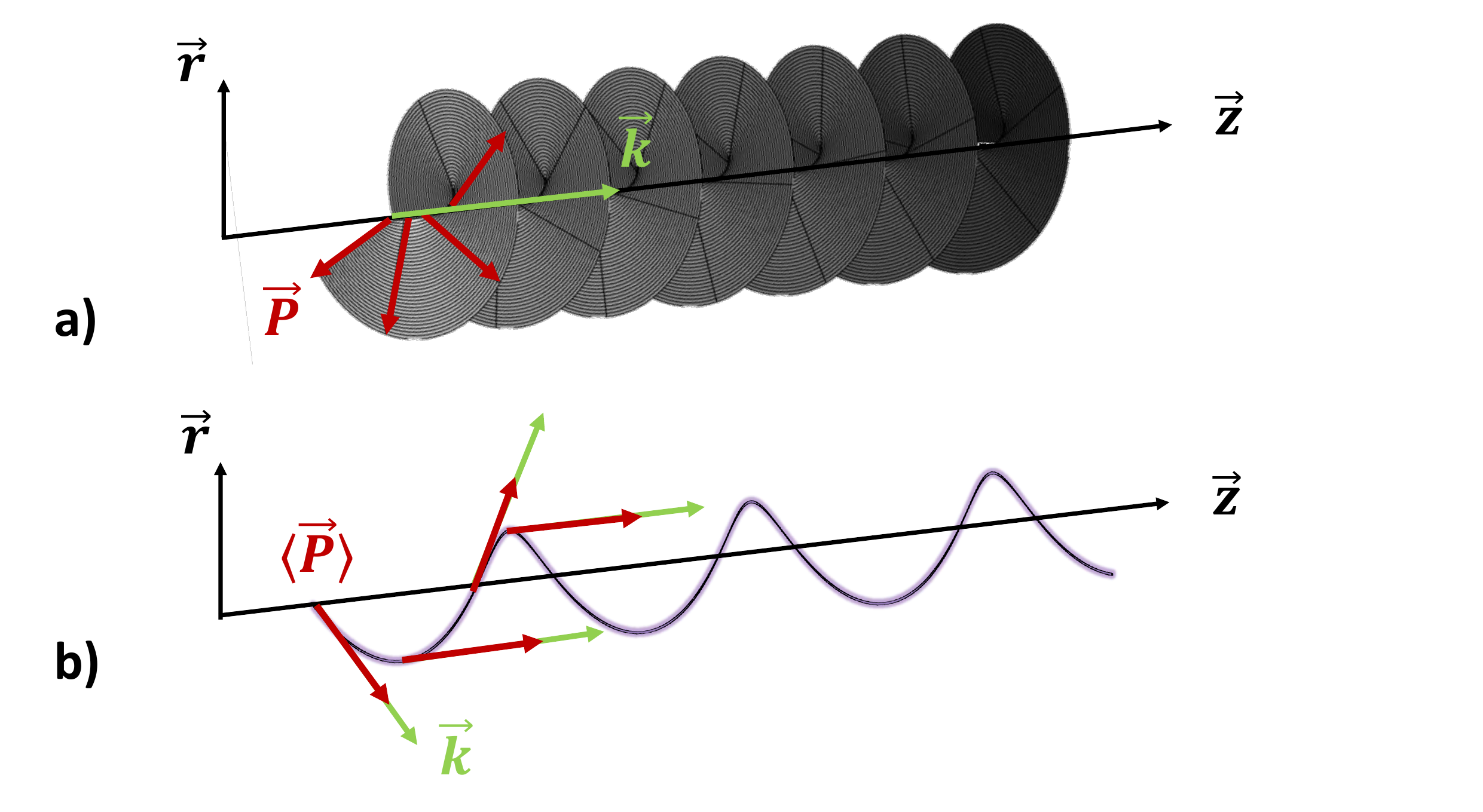}
	  \caption{Pictorial view of a) LOAM and b) TOAM beam propagation. 
	  }
	  \label{Fig0}
\end{figure}

Early studies of the LOAM of light by Allen et al. date back to more than thirty years ago\cite{allen1992orbital}. 
Over time, interest in LOAM beams has increased tremendously, thanks to their potential widespread applications. These range from telecommunications\cite{wang2012terabit,bozinovic2013terabit} to quantum optics\cite{fickler2016quantum}, holography\cite{heckenberg1992generation,oraizi2020generation}, and optical tweezers\cite{simpson1996optical,zhuang2004unraveling}, a research recognized by the award of the 2018 Nobel prize to Arthur Ashkin.
More recently, another approach of generating spiral beams via LOAM was proposed\cite{ustinov2021control}. It consists of giving light a power-exponent-phase $\phi=2\pi\ell(\varphi/2\pi)^n$, being $n$ a real number.
In all of these cases, the orbital angular momentum is parallel to the beam wave vector, and it is therefore dubbed longitudinal\cite{reviewOAM}: $\vec{L}_l = \hbar \ell \langle\vec{k}\rangle/k$. For a LOAM beam, the Poynting vector $\vec{P}$ periodically rotates around the wave vector, as depicted in  Fig.~\ref{Fig0}a.

An interesting, and somewhat less studied, case is that of beams possessing a transverse, as opposed to longitudinal, orbital angular momentum (TOAM)\cite{reviewOAM}. This type of beam can be more easily described when a ray optics approximation can be introduced, whereby we associate to the light beam a particle with a corresponding momentum $\langle\vec{P}\rangle$, which represents the average Poynting vector\cite{reviewOAM}. The resulting TOAM reads $\vec{L}_t=\vec{r}\times\langle\vec{P}\rangle$, where $\vec{r}$ is the position of the beam with respect to the origin of the reference frame. In this case, the average Poynting vector traces a helical path (see Fig.~\ref{Fig0}b). 

Recall that the total angular momentum of light ($\vec{J}$) can be expressed as the sum of two contributions: 
\begin{equation}
    \vec{J}=\vec{S}+\vec{L} = \vec{S}+\vec{L}_l+\vec{L}_t.
\end{equation}
Here $\vec{S}$ is the spin angular momentum (SAM) that, differently from the orbital momentum, characterizes the state of polarization of light. The SAM has attracted significant research interest, owing to its applications to nanophotonic Berry-phase devices, e.g., based on the spin Hall effect of light\cite{petersen2014chiral}.

Whereas LOAM beams have been mostly obtained so far by means of external optical components\cite{shen2019optical}, which poses limitations to their use in integrated optics\cite{forbes2016creation}, the generation of SAM and TOAM beams can be more easily achieved\cite{bliokh2013goos}. The latter, in particular, can be obtained by starting from circularly polarized beams (that carry SAM), thanks to the total angular momentum conservation in the refraction or reflection of light from either dielectric\cite{hosten2008observation} or metallic interfaces\cite{gorodetski2012weak}, where the spin-Hall effect of light (or Imbert–Fedorov shift\cite{bliokh2013goos}) occurs.

In this work, we propose and demonstrate how TOAM-carrying beams can be spontaneously generated in standard silica optical fibres, and seed spiral-shaped emission. As a cylindrical waveguide, an optical fibre is a natural beam shaper for TOAM beams. Our method consists of focusing in the fibre a laser beam with proper angles $\vartheta$ and $\varphi$, as sketched in Fig.~\ref{setup}a. This produces an azimuthal component to the beam wave vector, so that when propagating in the fibre, the beam starts twisting along the interface, eventually  converting (at least partially) the TOAM into a LOAM. As a result, an Archimedean spiral-shaped output light intensity is observed in the far-field, as shown in Fig.~\ref{setup}b.
By our approach, we can easily change the chirality of the spiral, by just inverting the sign of the incidence angle $\vartheta$, thus flipping the winding direction. 
In the nonlinear propagation regime, supercontinuum (SC) generation leads to high-brightness spectral broadening, covering all visible spectrum. We reveal that in the far-field, the SC is still a spiral beam, which now contains a mixture of different colours. Under proper input coupling conditions, these colours spontaneously separate in space, giving rise to a spiral-shaped rainbow emission. We conclude by pointing out analogies with the effect of conical emission in filamentation, that accompanies high-power laser beam propagation in air\cite{walter2010emission,walter2017emissions}.

\bigskip
\section*{Results}
\begin{figure}[!ht]
	  \centering
	  \includegraphics[width=0.8\textwidth]{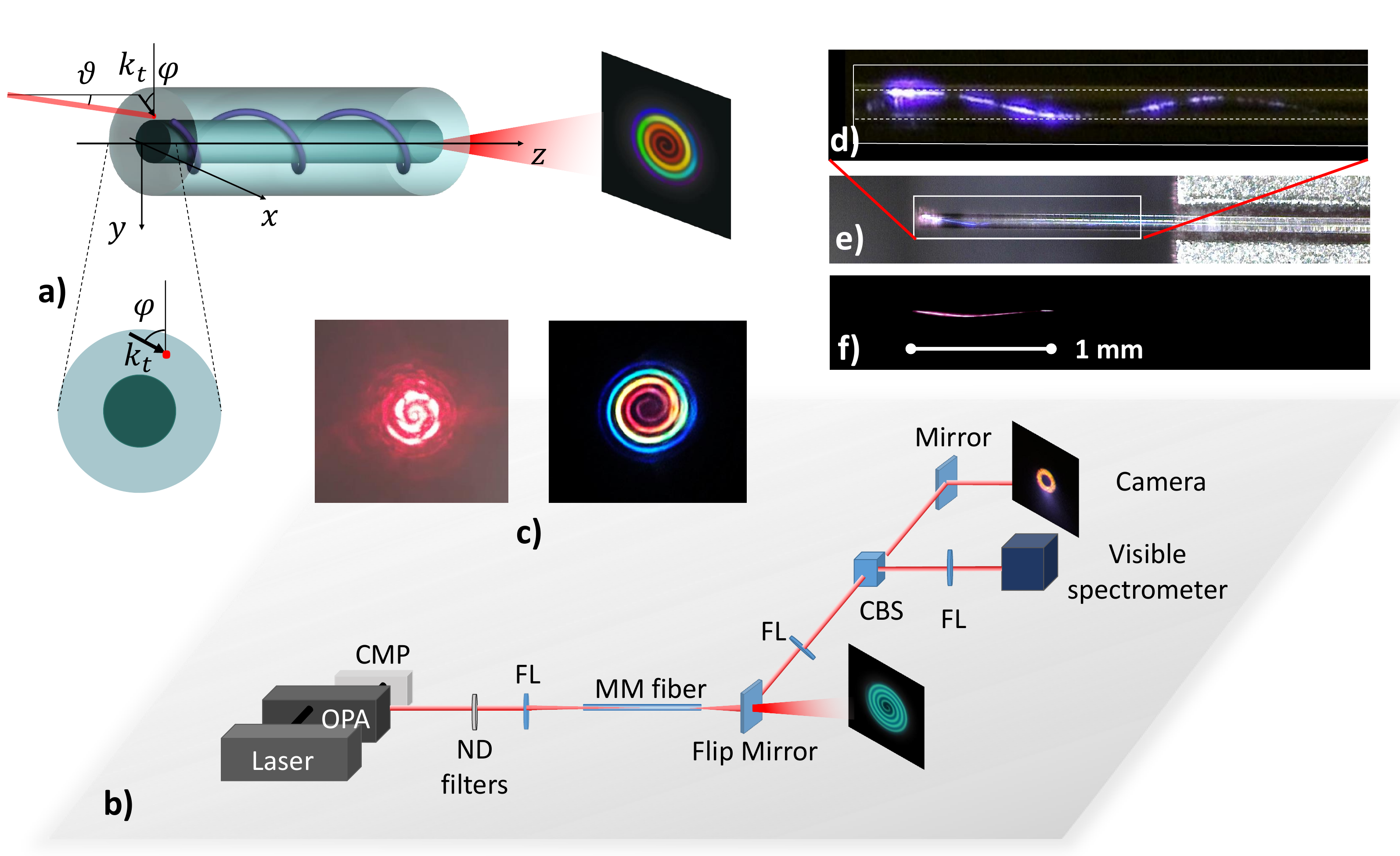}
	  \caption{(a) Schematic of fibre geometry and rainbow spiral emission; $\vartheta$ is the incidence tilt of the skew ray (or zenith angle), $\varphi$ is the azimuth angle, and $k_t$ is the projection of the laser beam wave vector on the fibre transverse plane. (b) Schematic of set-up where typical near- and far-fields are shown. The ND filter is used to vary input power, FL stands for focus lens, and CBS for cubic beam splitter. (c) Spiral emission from CW He-Ne laser (left) or from femtosecond pulse laser (right). (d) Zoom-in of (e), dashed lines indicate core-cladding boundaries. (e) Visualization of helical beam propagation in a graded-index (GRIN) fibre, thanks to upconversion luminescence from material defects and doping. Scattered white light on the right is due to the fibre holder. (f) Same as (d) but from step-index fibre.}
	  \label{setup}
\end{figure}

Our approach to generate spiral emission by means of commercial, standard telecom fibres is remarkably simple. We demonstrate that a laser beam, 
coupled with proper input zenith angle $\vartheta$ and azimuth angle $\varphi$ (see Fig.~\ref{setup}a) into the fibre, produces a spiral emission in the far-field, independently of its wavelength or polarization. The main conditions on the incidence beam angles are: small zenith angles (e.g., $\vartheta<2^\circ$), and azimuthal angles such that the projection of the incident beam wave vector on the transverse plane of the fibre ($k_t$) is tangent to the core/cladding interface, so that its radial component is vanishing (see Fig.~\ref{setup}a). 
The experimental set-up to study spiral emission is shown in Fig.~\ref{setup}b. As a source for SC generation, we used intense femtosecond laser pulses at the 1030 nm wavelength. As discussed in the Supplementary, we observed that spiral emission can be obtained with shorter wavelengths as well. As it exits the fibre, the beam is collected by an imaging lens to measure the near-field spatial profile and spectrum with a CCD camera and a spectrometer, respectively. By means of a flipping mirror we could also image the far-field.

The spiral emission shown in the left panel of Fig.~\ref{setup}c was obtained by means of a He-Ne CW laser. On the other hand, the true-colour picture on the right panel of the same figure was obtained by matching four different conditions: TOAM seeding, spiral emission, SC generation, and colour spatial separation. In the following, we investigate all of these elements, one at a time. 

\subsection{TOAM seeding.}
In our experiments, we removed the external plastic coating of the fibre, so that we can directly track the beam propagation inside the fibre by naked eye, by exploiting the phenomenon of upconversion luminescence due to the presence of doping, and of intrinsic and extrinsic defects in MMFs\cite{hansson2020nonlinear,ManginiFerraro} (see Methods). We underline that TOAM-carrying beam propagation is a linear phenomenon. High-power pulses are only used in order to excite the luminescence, which is needed for tracking the path of light beams.
As illustrated in Fig.~\ref{setup}, a helical beam path in MMFs was observed when injecting, with appropriate incidence angles, the laser beam in proximity of either the core-cladding or the cladding-air interfaces, respectively. Results in Fig.~\ref{setup} refer to both step-index (SI) and graded-index (GRIN) MMFs (in the Supplementary material we also report the case of a singlemode fiber). Fig.~\ref{setup}d-f shows beam paths as revealed by the multiphoton-absorption generated luminescence, which is measured by a microscope positioned on top of the MMFs. 

With either GRIN or SI fibres, we may always clearly recognize a visible photoluminescence, that traces helical beam paths. These can be associated with the propagation of TOAM beams, which are internally twisting around the core-cladding or the cladding-air interfaces, respectively. It is interesting to observe the difference in colour between luminescence from either a GRIN fiber (Fig.~\ref{setup}d-e) or a SI fiber (Fig.~\ref{setup}f). This is due to the presence of doping in the core of a GRIN fibre, which produces a blue luminescence. Whereas silica intrinsic defects produce a red light at the core-cladding interface of a SI fibre\cite{ManginiFerraro}. For the same reason, beams tracing a helical path around the cladding-air interface shine red light from both SI and GRIN fibres.
The propagation of TOAM-carrying beams with helical paths was previously reported in a glass rod\cite{bliokh2008geometrodynamics}: it was shown that key to produce a TOAM beam is the presence of a refractive index gradient at the glass-air interface. Accordingly, we also found that the generated luminescence only traces a helical beam path when the input beam touches the inner or the outer boundaries of the fibre cladding.

\begin{figure}[!ht]
	  \centering
	  \includegraphics[width=0.85\textwidth]{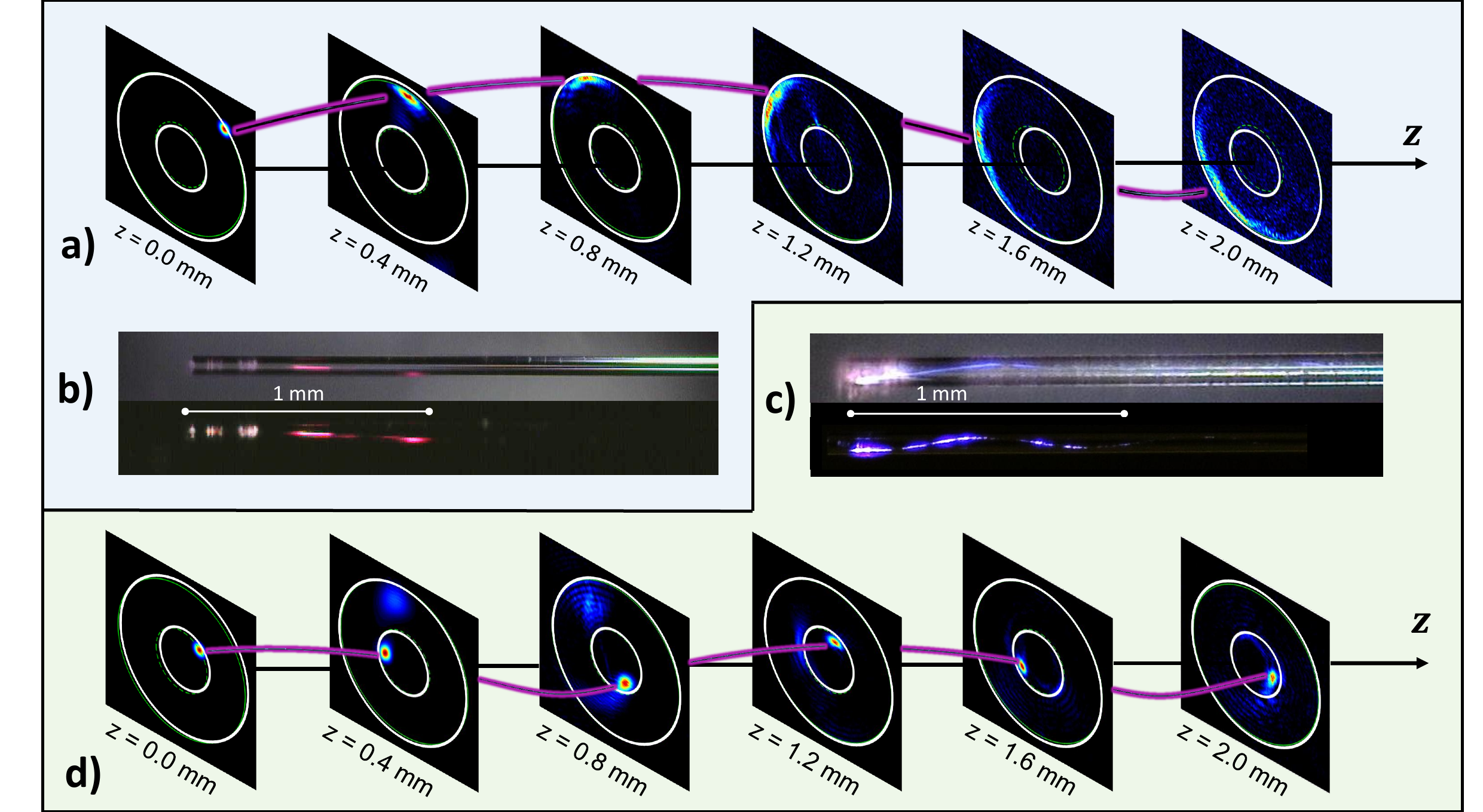}
	  \caption{TOAM carrying beam tracking, obtained when the input beam is incident at the cladding-air (a,b) or the core-cladding interface (c,d), respectively. The purple curves in (a) and (d) represent the Poynting vector trajectory. Simulated near-field intensities are compared with experimentally observed luminescence intensities. 
	  }
	  \label{nuova}
\end{figure}

In order to confirm the observed TOAM carrying beam trajectories, we performed a series of numerical simulations (see Methods for details). In Fig.~\ref{nuova}a,b, we illustrate the case of a beam generated at the cladding-air interface. For a comparison, the case of core-cladding interface incidence is also reported in Fig.~\ref{nuova}c,d.
In both cases, a good agreement is found between numerical and experimental values of the winding period, which grows linearly proportional to the curvature radius. To visually compare numerical and experimental results, we highlight the Poynting vector trajectory with a purple curve (similar to the one in Fig.\ref{Fig0}b) crossing the numerical simulation images at the maximum intensity points. Simulations also show that, owing to diffraction, the beam size progressively widens while internally twisting around the interface, which eventually invalidates the ray-optics approximation. Nevertheless, the total angular momentum conservation law ($\vec{J}=const.$) ensures that the TOAM beam maintains its angular momentum during the propagation\cite{reviewOAM}. In the next Section, we will show how the TOAM is converted into a power-exponent-phase LOAM during the propagation, so that, under suitable conditions, the beam acquires a proper amplitude and phase, in order produce an Archimedean  spiral shaped intensity profile in the far-field.
\subsection{Spiral emission.}
When we observe spiral emission, the luminescence always traces a helical shape. Therefore, we can state that the spiral emission is seeded by the TOAM. In other words, generating a spiral shape in the far-field intensity requires a symmetry breaking of the propagating field (e.g. in the phase) in the plane transverse to its propagation direction, which always produces a TOAM \cite{bliokh2013goos,zhao2017identifying}. Indeed, such spiral emission from multimode optical fibers was earlier observed, and explained by means of a ray tracing model \cite{coutts2018diagnostic}. In that work, a helical trajectory for the propagating beam is imposed by a controlled microbending along the fiber. Whereas, in our case, the TOAM is provided by the input coupling conditions.
Remarkably, our spiral emission generation method is linear, hence it does not require particularly high laser powers. As detailed in the Supplementary, spiral emission could be generated independently of the laser average power and pulse duration. For example, we could generate a red spiral beam by using a CW He-Ne laser (see Fig.~\ref{setup}b-left), and even by using a standard laser pointer as a source (see Supplementary)! Furthermore, spiral emission does not appear to be affected by varying the input state of polarization, and it is wavelength-independent.

\begin{figure}[!ht]
	  \centering
	  \includegraphics[width=1\textwidth]{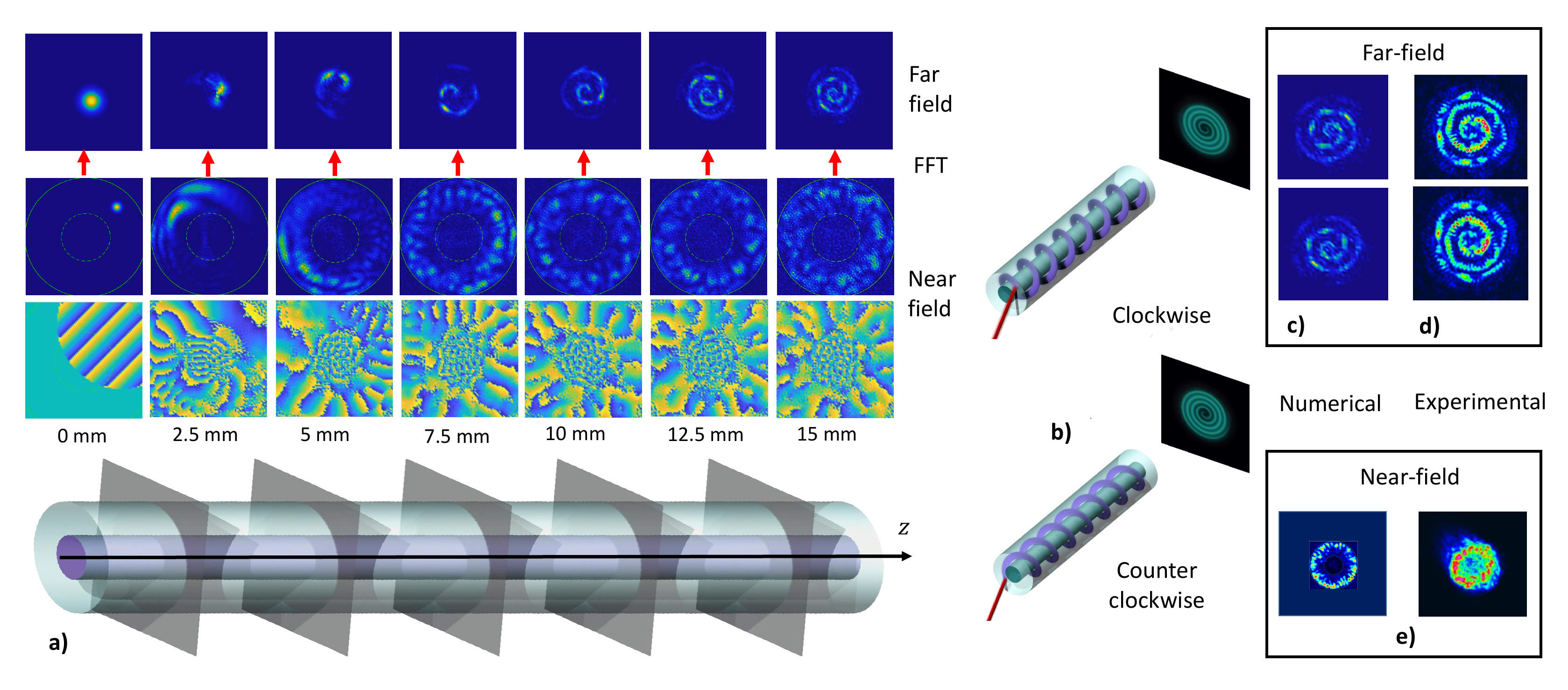}
	  \caption{(a) Sketch of the first 1.5 cm of the fibre (bottom);  The near-field (amplitude and phase) is numerically computed each 2.5 mm, corresponding to the position where the transverse planes cut the fibre (centre). The far-field is calculated as the Fourier transform of the near-field, showing spiral emission formation (top). (b) Tunability of optical chirality. The spiral emission TC sign depends on the input beam position with respect to the fibre centre. (c) Numerical confirmation of optical chirality inversion. (d) Experimental results corresponding to (c). (e) Comparison of numerical and experimental near-field intensities. All simulations and experiments were performed with an input peak power of 3 MW, a wavelength of 1030 nm, $\vartheta = 1.5 ^\circ$ and $\varphi = 45 ^ \circ$.
	  }
	  \label{numerical}
\end{figure}

In order to theoretically reproduce the observed spiral emission, we performed extensive numerical simulations. 
Fig.~\ref{numerical}a shows that a beam, which is initially focused on small spot inside the fibre cladding, progressively occupies all of the cladding area, and it acquires a phase profile that rapidly varies in the azimuthal direction (see Methods for details of the numerical model). In Fig.~\ref{numerical}a we also plot the far-field, which would be generated by propagating the spatial field distribution at given specific points inside the fibre. Specifically, the far-field is calculated as the Fourier transform of the near-field in each slice. In the Supplementary materials, we provide videos of amplitude and phase evolution of both the near- and far-fields along the beam propagation. 
Thanks to the total orbital angular momentum conservation, the initial TOAM is converted into a LOAM. In particular, as Fig.\ref{numerical}a shows, the acquired phase nonlinearly scales with the angle $\varphi$. The obtained phase pattern at 5 mm of propagation is pretty similar to that of Ref.\cite{ustinov2021control}.

Simulations also explain how to tune the chirality of the spiral. Depending on the beam input position, one can reverse the chirality, from clockwise to counterclockwise. This is illustrated in Fig.~\ref{numerical}b. Diametrically opposed input positions correspond to opposite winding, hence flipping the chirality sign. Excellent qualitative agreement between simulations and experiments was achieved, as shown in Fig.~\ref{numerical}c-d and Fig.~\ref{numerical}e for the far- and near-field respectively. 

As shown in Fig.~\ref{numerical}, the TOAM beam intensity at the fibre output facet has an annular shape, similar to the case of LOAM beams. Note that the generation of the so-called hollow beams, with a helical path around the core induced by tilted laser-fiber coupling, was previously reported\cite{schweiger2010generation}. However, it remained unnoticed that, besides the annular transverse intensity beam pattern, a proper phase pattern (associated with TOAM) is necessary, in order to generate spiral emission in the far-field. 
Therefore, the conditions for spiral emission are stricter than those required for the formation of an hollow beam.
While the latter can be realized with relatively large input tilt angles and position offsets\cite{schweiger2010generation}, spiral emission is only obtained when the input laser beam is focused close to fibre index interfaces, and with a small tilt angle $\vartheta$ (between 1 and 2 degrees). We observed, both numerically and experimentally, that if the incidence is not grazing, the beam never reaches the proper amplitude and phase which are necessary to generate a spiral pattern in the output far-field intensity profile. Specifically, we noted that as $\vartheta$ increases, the range of spiral emission existence tends to decrease.

Another critical condition for spiral emission involves the fibre length. One can appreciate from Fig.~\ref{numerical}a that spiral formation requires a certain minimum propagation length (estimated to be of a few millimeters with our parameters). At the same time, the fire must be short enough, in order to avoid substantial beam spreading inside the core, thus washing out the annular shape of the near-field intensity profile.
We estimated, both numerically and experimentally, the optimal threshold length to be about 2 cm for $\vartheta \simeq 2^\circ$.
With such short fibre lengths, we managed to keep the output transmission above 90\% even at MW input powers
\cite{zitelli2020high}. 
\subsection{SC generation.}
Whenever the injected power of a near-infrared femtosecond laser source was high enough to produce SC generation, i.e., a wide nonlinear spectral broadening of the input laser pulses, a coloured spiral intensity profile in the far-field was emitted from the fibre output facet. 
In Fig.~\ref{colour_separation}b,c we report examples of spectra 
obtained from SI and GRIN MMFs, respectively, when varying the input peak power. When the latter reaches 48 MW, the whole visible spectral range is covered, producing all of the images shown in Fig.~\ref{colour_separation}d. 
Due to the absence of clear peaks in the visible range, and because of the nearly symmetric broadening of the pump spectra, we may ascribe the observed SC mechanism to self-phase modulation. This must occur radially, in order to explain the rainbow spatial distribution of colors. 
Unfortunately, our numerical model diverges at peak power values well below those necessary to generate a SC covering the whole visible spectral range. Nevertheless, as an alternative approach to explain the origin of colored spiral emission, we simulated the separate propagation of a low peak-power pulse, with different carrier wavelengths across the visible spectral range. As a matter of fact, a colour-separated spiral can be reproduced by the incoherent summation of far-field intensities, independently generated by different pulses with separate wavelengths, as depicted in Fig.~\ref{colour_separation}a. For all spectral components, the fiber length is kept fixed to 1.5 cm, the position offset is 60 $\mu$m, the beam $1/e^2$ diameter is 11.9 $\mu$m, $\vartheta=1.5^\circ$,$\varphi=45^\circ$, the input peak power and pulsewidth are 1 kW and 180 fs, respectively. Furthermore, chromatic dispersion is induced by the following wavelength dependence of the cladding refractive index: $n(350nm) = 1.4787$, $n(400nm) = 1.4701$, $n(500nm) = 1.4623$, $n(550nm) = 1.4599$ and $n(650nm) = 1.4565$.

\subsection{Color separation.}
\begin{figure}[!ht]
	  \centering
	  \includegraphics[width=0.9\textwidth]{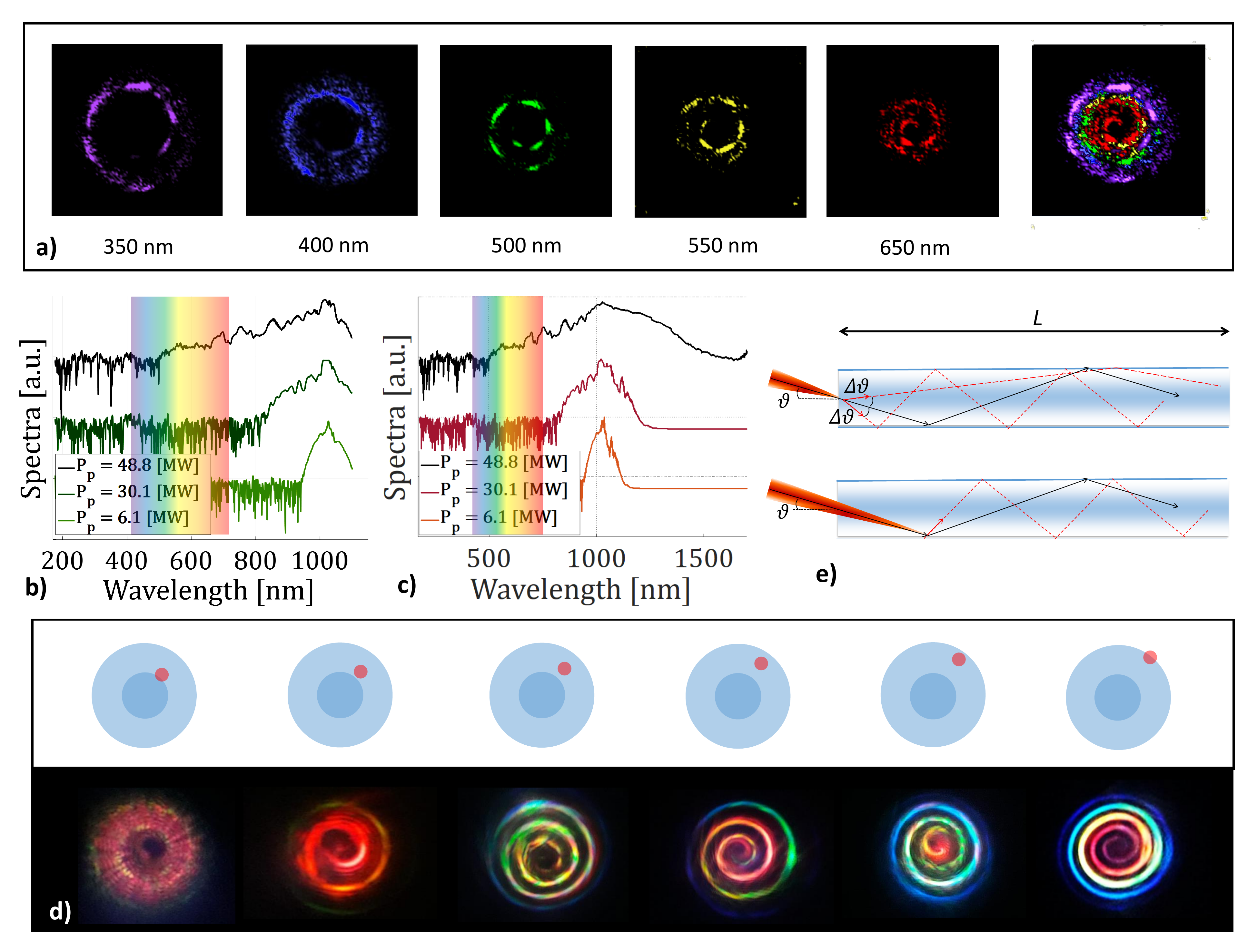}
	  \caption{(a) Rainbow spiral simulation as a sum of the single color component. (b-c) Fibre output spectra at different input peak powers (up to 48 MW) at 1030 nm in step- and graded-index fibres respectively keeping the same coupling conditions. (d) Far-field image of graded-index fibre when varying the input beam position keeping the peak power at 48 MW, while keeping $\vartheta = 1.5 ^\circ$ and $\varphi = 45 ^ \circ$. (e) Ray optics sketch of conical emission analogy. On the top, the SC is generated inside the fiber volume. On the bottom, the case of SC generates at the air-cladding interface.}
	  \label{colour_separation}
\end{figure}

\begin{figure}
	  \centering
	  \includegraphics[width=0.7\textwidth]{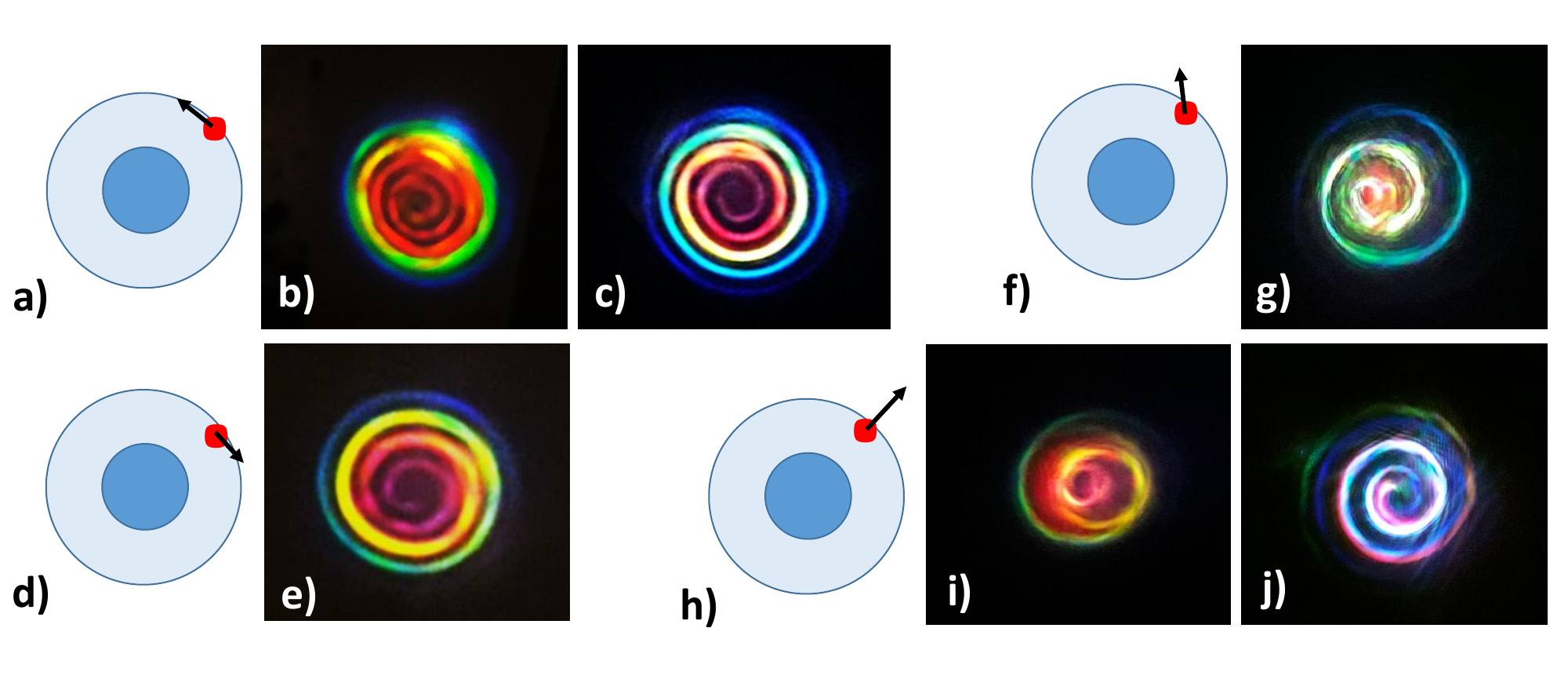}
	  \caption{(a) Sketch of the input configuration for rainbow spiral emissions obtained in either SI (b) or GRIN (c) fibres. Black arrows represent the transverse plane component of the input wave vector. (d) Symmetric situation with respect to (a), leading to flipping the topological charge which results in a change of chirality, as seen in (e). (f-g) Superposition of two spirals with opposite chiralities and different intensities (g) (heart-like spiral emission), when the transverse wave vector has a nonzero radial component (f). (h-i) Same as in (f-g), when the transverse wave vector points along the radial direction. (j) Bistable case, resulting in two spirals beam with same chiralities. For all pictures shown, the input power and wavelength were 48 MW and 1030 nm, respectively and the fibres were 2 cm long. All the images but (b) refer to GRIN fibres.}
	  \label{change_TC}
\end{figure}

Finally, we could observe that colour spatial separation within the spiral only depends on the input coupling condition, i.e., the position of the incident laser beam on the fibre, while keeping the same spectral composition. In Fig.~\ref{colour_separation}d we show that speckled features appear when the input beam is only partially focused inside the core. By gradually moving deeper into the cladding, spiral emission becomes clearer, and it is characterized by a disordered mix of colours. Finally, when the beam crosses the cladding-air interface, we obtain the rainbow spiral emission as shown in Fig.~\ref{setup}b-right. In our experiments, multiple arm spirals were also observed. An overview of these hybrid order spiral beams is presented in Fig.~\ref{change_TC}.

During the experiments, we observed that also the shape of the spiral emission strictly depends on the specific input coupling conditions. The perfect spiral shape configuration is obtained when the input beam wave vector radial component is vanishing. In Fig.~\ref{change_TC}a we report the particular case of a beam focused on the cladding-air interface: here the black arrow indicates the in-plane component of the wave vector. This configuration gives rise to a rainbow spiral emission in both SI and GRIN fibres, as illustrated in Fig.~\ref{change_TC}b and c, respectively. If one reverses the direction of the arrow (i.e., by rotating the azimuthal angle), the chirality of the spiral flips its sign (see Fig.~\ref{change_TC}d-e).

However, one can have intermediate configurations, where the radial component of the wave vector is nonzero. We show in Fig.~\ref{change_TC}f-g an example of input coupling condition, for which the beam appears as a superposition of spirals with opposite chiralities. One can remark the formation of a heart shape at the centre of the far-field image. In the extreme case, where the transverse component of the wave vector is directed along the radius (Fig.~\ref{change_TC}h), the two counter-rotating spirals have exactly the same intensity, producing the far-field which is shown in (Fig.~\ref{change_TC}i). This configuration is bistable, and in the presence of small fluctuations, one may observe the generation of highly complex far-field intensity patterns. Finally, we report in Fig.~\ref{change_TC}j the case of a two-arms-spiral emission.

\bigskip
\section*{Discussion}
The observed Archimedean spiral emission can be explained by the help of the analogy with the generation of angular momentum light beams from single filaments in air\cite{walter2010emission,walter2017emissions}. This occurs when an intense laser pulse increases the refractive index of air, owing to the Kerr effect. As a consequence, the pulse collapses, and a plasma channel is generated along the optical axis. When beam self-focusing and plasma defocusing compensate for each other, the pulse transforms into a filament, surrounded by an energy reservoir. This is reminiscent of light propagating in optical fibres: the plasma channel acts as the core, and the energy reservoir as the cladding\cite{walter2017emissions}. 
Upon propagation in the filament, a portion of radiation is lost, being emitted at specific angles, a process which is well known as conical emission\cite{golub1990optical, nibbering1996conical, kosareva1997conical,  di2003spontaneously, conti2003nonlinear, faccio2007spatio, aumiler2005femtosecond}. 
Recently, Walter at al. demonstrated angular momentum beam generation by means of deformable mirrors, which permitted to transform a conical emission into spiral emission \cite{walter2010emission,walter2017emissions}. 
The latter was ascribed to the difference of group velocity between the plasma channel and the energy reservoir, which allows for a continuous interaction between their modes. The background energy flows helically around the plasma channel (just as seen in Fig.~\ref{setup}c), thus slightly deviating the filament, and impressing a helical path to the beam. In fibres, nonlinearity is not necessary to obtain an optical angular momentum, because the cladding (reservoir) and the core (filament) are defined by the linear index profile. Moreover, no external spatial modulation is needed, since the cylindrical fiber geometry spontaneously creates a TOAM, when symmetry-breaking is seeded by a small tilt of the angle of transverse incidence. 

The analogy with radiation emission by single filaments in air allows for an alternative explanation of the mechanism of colour separation in rainbow SC spiral emission from MMFs. In conical emission accompanied by SC generation, shorter wavelengths are emitted with wider angles.
The wavelength spread produced by source broadening depends on the Kerr effect-induced electron density gradient: the larger the gradient, the more shifted the generated wavelengths. When SC is generated, its wave vector has a radial component whose modulus linearly grows with optical frequency. This results in a narrower (larger) angle of emission for long (short) wavelengths. In particular, since the radial electron density gradient varies continuously from zero to a maximum value, rainbow rings are generated. This phenomenon has been attributed to self-phase modulation in the radial direction\cite{couairon2007femtosecond,smetanina2010supercontinuum,kandidov2011formation}. 

When applied to our case, the mathematical model to describe the frequency- angular intensity distribution of spectral components $S_{an}(\vartheta,\lambda,z)$ is\cite{kandidov2011formation}
\begin{equation}
    S_{an}(\vartheta,\lambda,z)=S_{0}(\vartheta,\lambda,z)\ell(z)^2
    \text{sinc}{\frac{\Delta\vartheta}{2}},
\end{equation}
where $\Delta\vartheta$ is the phase excursion of radiation from a broadband point source with distribution $S_0$ and propagating with group velocity $v_g$ over distance $\ell(z)$ which reads as\cite{kandidov2011formation}
\begin{equation}
    \Delta\vartheta=\frac{2\pi\ell(z)}{\lambda_0}\bigg[\bigg(1-\frac{\lambda_0}{\lambda}\bigg)\frac{c_0}{v_g}-\bigg(1-\frac{\lambda_0n(\lambda)}{\lambda n_0}\cos(\vartheta)\bigg)n_0\bigg].
\end{equation}
\noindent Here $n_0$ is the refractive index, $n(\lambda)$ is material dispersion in silica, and $c_0$ the speed of light in vacuum.
Since we spontaneously generate a TOAM beam, the helical path of the photon now leads to rainbow-like spiral emission, instead of a conical emission. The filamentation analogy also helps to explain the colour separation shown in Fig.~\ref{colour_separation}d. Whenever SC generation takes place inside the cladding volume, the pump propagation direction is locally a symmetry axis, and two wavevectors are associated with the same wavelength, which is equivalent to conical emission. This configuration is sketched on the top of Fig.~\ref{colour_separation}e by an intuitive ray-optics picture. The initial spread $2 \Delta \theta$ gets wider after each reflection, leading to interference which is responsible for colour mixing in the far-field. One can estimate that very small emission angles ($\Delta\theta < 10^{-5} rad$) are needed in order to incur into interference. Things drastically change when the input beam is focused right at the cladding-air interface (bottom of Fig.~\ref{colour_separation}e). This is because the cladding edge breaks the symmetry along the beam propagation, and a one-to-one correspondence between wavelength and wavevector is imposed by the  electromagnetic field continuity condition at the boundary. The resulting lonely ray does not suffer from any interference, and a rainbow colour distribution is observed as in the case of conical emission in air.

\bigskip
\section*{Conclusion}
We observed the generation of multicolour spiral beam by using standard commercial silica optical fibres. The main trick consists of choosing appropriate fibre coupling conditions, in order to seed a TOAM that, combined with the self-focusing process, permits to generate a large spectral broadening. 
Our method of spiral beam generation has several advantages. First of all, employing optical fibres allows for easier integration with existing devices and, while conventional LOAM beam generation methods rely on spatial light modulators, here we exploit the conversion of the TOAM, that is intrinsically provided by the cylindrical geometry of the fibre. Moreover, the method is quite robust: it does not depend on the source state polarization, power, and wavelength. The spiral emission formation in fibres is analogous to light filamentation in air. Accordingly, in analogy with conical emission, we could explain the observed rainbow spiral emission in MMFs.
Our results provide a substantial advance in structured light generation, and pave the way to the practical application of TOAM beams in data storage, super-resolution, and nanoscale microscopy technologies. Moreover, 
thanks to the low sensitivity to environmental perturbations with respect to in-air LOAM beam generation, our method can be applied to implement immersible optical tweezers for applications in biology as well.
\bigskip
\begin{methods}
\subsection{Experimental Setup}
Our pulsed light source was a Yb-based laser (Light Conversion PHAROS-SP-HP), generating pulses of 180 fs with 1-100 kHz repetition rate. The laser beam was focused on the input facet of the fibre, and incident with controllable tilt angles $\vartheta$ and $\varphi$ with respect to the fibre axis $z$, and with a beam diameter of about 10 ${\mu}$m at $1/e^2$ of peak intensity. Both the parabolic GRIN fibre and the step-index fibre had a core radius $r_\text{c}$=25 $\mu$m, cladding radius 62.5 $\mu$m, and a cladding index $n_\text{clad}$=1.45 at $\lambda$ = 1030 nm.
The relative core-cladding index difference was $\Delta$=0.0103 or 0.0120 for the GRIN and the step-index fibre, respectively. The index grading is realized by Germanium doping.
The top view images in Fig.~\ref{setup}c-e were taken with a Dinolight2.0 digital microscope for 14~mW average input power from the femtosecond laser at 1030 nm, at the repetition rate of 2~kHz. Spectra were collected by a miniature fibre optics spectrometer (Ocean Optics USB2000+), with 200-1100 nm spectral range and an optical spectrum analyzer (OSA) (Yokogawa AQ6370D), while far and near-field images were taken by a Reflex digital camera (Nikon D850) and a Gentec Beamage-4M-IR CCD camera, respectively. Input and output average powers were measured by a thermopile power meter (GENTEC XLP12-3S-VP-INT-D0).
Our CW light source was a 4 mW HeNe-based laser (Thorlabs HNLS008L-EC).

\subsection{Numerical model}
The numerical model developed to describe the process of spiral emission generation in MMFs is the $(3D+1)$ GNLSE (or Gross-Pitaevskii equation\cite{pitaevskii}) in its vectorial form, involving a single field for each polarization, including all frequencies and modes\cite{zitelli2020high}: 
\begin{align}\label{3d+1}
&\frac{\partial A_p(x,y,z,t)}{\partial z}=\frac{i}{2k} \bigg(\frac{\partial^2A_p}{\partial x^2}+\frac{\partial^2 A_p}{\partial y^2}\bigg)-i\frac{\beta_2}{2}\frac{\partial^2A_p}{\partial t^2}+ \frac{\beta_3}{6}\frac{\partial^3A_p}{\partial t^3}+ i\frac{\beta_4}{24}\frac{\partial^4A_p}{\partial t^4} -\frac{\alpha}{2}A_p+\\ \nonumber
+&i\frac{k}{2}\bigg[\frac{n^2(x,y)}{n^2_{0}}-1\bigg]A_p+
i\gamma \bigg(1+iK_2+\frac{i}{\omega_0}\frac{\partial}{\partial t}\bigg)\bigg[(1-f_R)A_p\bigg(|A_p|^2+\frac{2}{3}|A_q|^2+\frac{1}{3}A_p^2e^{-2i\omega_0t}\bigg)+\\ \nonumber
+& f_RA_p\int_{-\infty}^t d\tau h_R(\tau)\bigg(|A_p(t-\tau)|^2+\frac{2}{3}|A_q(t-\tau)|^2\bigg)\bigg]
\end{align}
with $k=n_0\frac{2\pi}{\lambda}$, $\gamma=n_2\frac{2\pi}{\lambda}$, $K_2=\frac{\alpha_2}{2\gamma}$, and $f_R=0.18$.
In Eq.~(\ref{3d+1}), the two polarizations $p,q=x,y$ are nonlinearly coupled. Terms in the right-hand side of Eq.~(\ref{3d+1}) account for: transverse diffraction, second, third, and fourth-order dispersion, linear loss, the waveguiding term with refractive index profile $n(x,y)$ and core index $n_\text{0}$, Kerr and Raman nonlinearities (with nonlinear coefficient $\gamma$  and fraction  $f_R$), respectively. In Eqs.~(\ref{3d+1}) we neglect the contribution of polarization mode dispersion: we numerically verified that its effects are negligible for the short fibre lengths (few centimeters) involved in our experiments. Nonlinearities include self-steepening, third-harmonic generation (THG), and two-photon absorption (TPA) (with coefficient $\alpha_2$).

In simulations, we used the following GRIN fibre parameters: core radius, cladding radius, and relative index difference are already provided in the experimental setup subsection; dispersion parameters are $\beta_2=18.9\, \text{ps}^2/\text{km}$ at 1030 nm, $\beta_3=0.041\,\text{ps}^3/\text{km}$, $\beta_4=-5.3\times 10^{-5}\, \text{ps}^4/\text{km}$; nonlinear parameters are $n_2=2.7\times10^{-20}\, \text{m}^2/\text{W}$, $\alpha_2=1\times10^{-16}\, \text{m}/\text{W}$; $h_R(\tau)$ with the typical response times of 12.2 and 32 fs, respectively\cite{Stolen:89,Agrawal01}. We included the wavelength dependence of the linear loss coefficient $\alpha$, as reported for standard SM glass fibres\cite{shubert}.
The input beam was modeled as a Gaussian beam with $w_0=5$ $\mu$m waist (10 $\mu$m diameter); we used a Gaussian temporal shape with full-width-at-half-maximum (FWHM) pulse-width $T_\text{FWHM}=180$ fs. The input beam was coupled at different points of the cladding or the core, and tilted by small angles (up to 15 degrees).

\end{methods}

\bigskip
\section*{Bibliography}
\bigskip
\bigskip
\bigskip
\bibliography{biblio}


\begin{addendum}
 \item This work has received funding from the European Union Horizon 2020 research and innovation program under the European Research Council (Grants No. 740355 and No. 874596), the Italian Ministry of University and Research (Grant No. R18SPB8227), and the Marie Skłodowska-Curie Grant No. 713694 (MULTIPLY). A.T., V.C., and T. M. acknowledge the financial support provided by: the French ANR through the “TRAFIC project: ANR-18-CE080016-01”; the CILAS Company (ArianeGroup) through the shared X-LAS laboratory; the “Région Nouvelle Aquitaine” through the projects F2MH and Nematum; the National Research Agency under the Investments for the future program with the reference ANR-10-LABX-0074-01 Sigma-LIM.
\\
\newline
Author contributions\\
F.M and M.F. carried out the experiments, M.Z. developed and performed the numerical simulations. All authors analysed the obtained results, and participated in the discussions and in the writing of the manuscript.\\
\\
The authors declare that they have no
competing financial interests.\\
Correspondence and requests for materials
should be addressed to \\Fabio Mangini~(email: fabio.mangini@unibs.it).
\end{addendum}
\input{Supplementary}

\end{document}

%% file: Supplementary.tex

\section*{Supplementary}

\bigskip
\subsection{Spiral emission as a function of input pulse duration, peak power, and wavelength.} 
\begin{figure}
	  \centering
	  \includegraphics[width=0.8\textwidth]{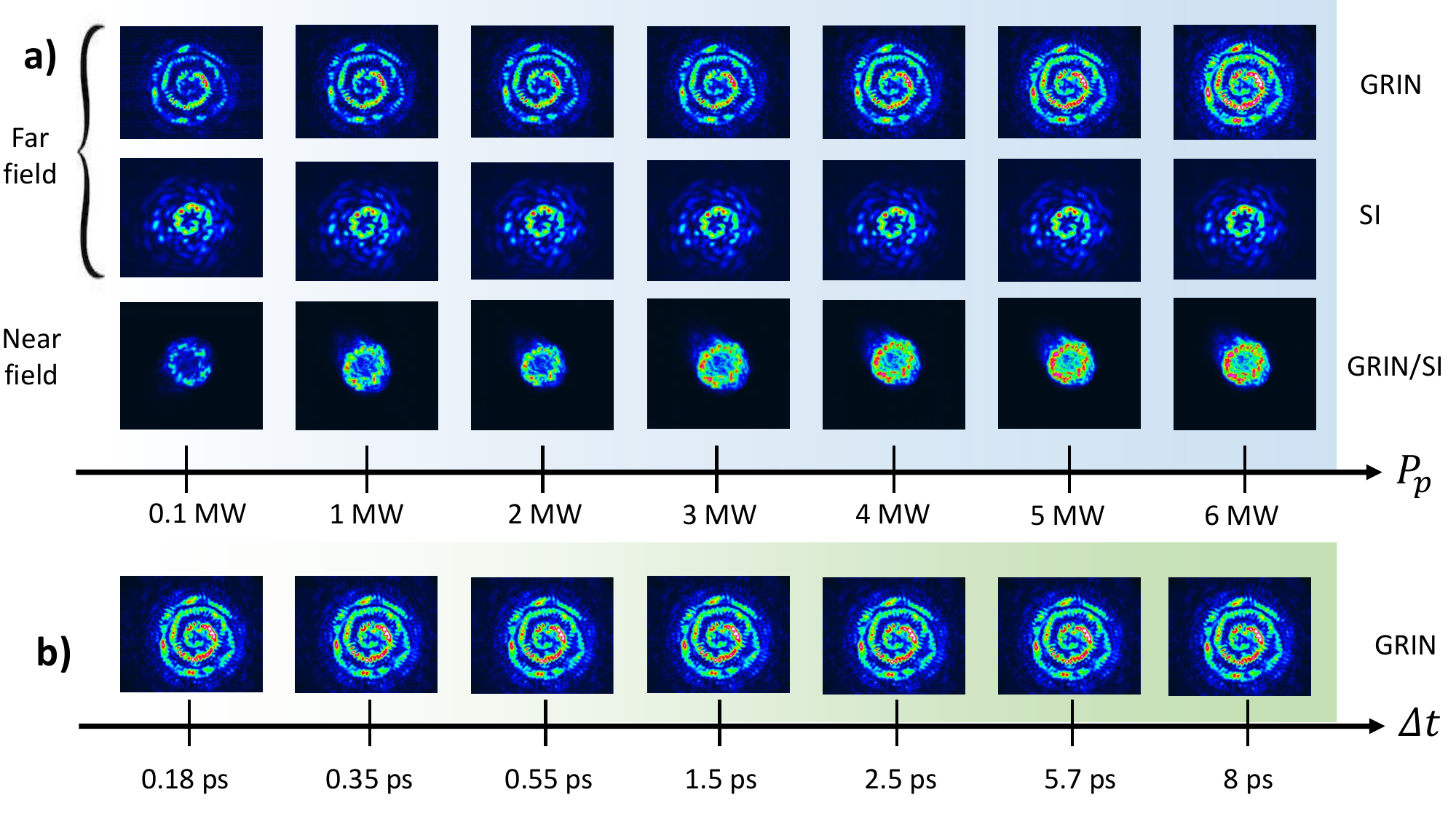}
	  \caption{(a) Near and far-field images at the output of a 2 cm GRIN or SI fibre. The input peak power varies from 0.1 MW to 6 MW, with a repetition rate 50 kHz. The source wavelength is 1030 nm, while the fibres are 1.5 cm long. The coupling angles are $\vartheta=2^{\circ}$ and $\varphi=45^{\circ}$. (b) Far-field images from GRIN fibre, under the same conditions as (a). Here the input average power is maintained at 0.1 mW, while the pulse duration varies between 0.18 to 8 ps}
	  \label{change_parameters}
\end{figure}

The objective of this section is to show that spiral emission by injecting a laser beam in an optical fibre is not affected by a change of different parameters of the source beam. We demonstrate that, by showing images of the far-field at the fibre output, when we vary either the peak power, the time duration of the incident laser pulse, or the source wavelength.
Fig.~\ref{change_parameters}a and b show far-field (first two rows) and near-field images (third row) at the output of 2 cm long GRIN or SI fibres. The near-field is only reported for the GRIN fibre case. Here the wavelength and the pulse duration are kept at 1030 nm and 180 fs, respectively, while the input peak power is varied between 0.1 MW and 6 MW. These values remain below the SC generation threshold, so that spiral emission is only observed in the near-IR (i.e., at the pump wavelength). By comparing the different images, we can see that that only intensity variations occur (owing to the increase of input power), while the spiral shape remains unchanged.
Fig.~\ref{change_parameters}b shows the evolution of the far-field intensity from the output of a 2 cm length of GRIN MMF at 1030 nm, with 0.1 mW of input average power, when the pulse duration is varied from 180 ps to 8 ps. As we can see, spiral generation is independent of the time duration of the input pulse as well, which confirms the linear nature of the phenomenon.

\begin{figure}
	  \centering
	  \includegraphics[width=\textwidth]{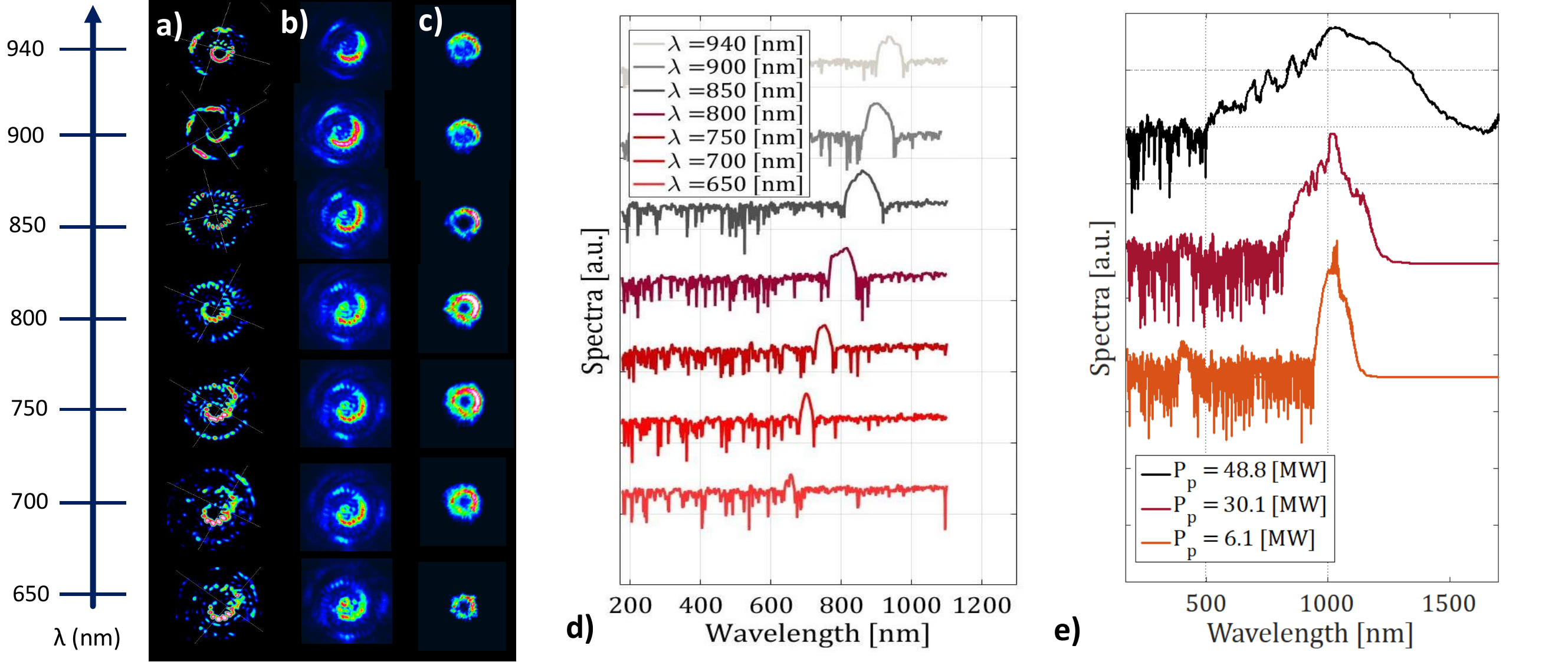}
	  \caption{(a) Far-field images from a 2 cm long GRIN when the wavelength of the laser source varies between 650 and 940 nm. The input power is kept at 0.1 MW, with a 50 kHz repetition rate; the input beam coupling angles are $\vartheta=2^{\circ}$ and $\varphi=45^{\circ}$.
	  (b) Same as in (a), for a SI fibre. (c-d) Near-field output intensity and spectra corresponding to (a). (e) fibre output spectra at different input peak powers (up to 48 MW) at 1030 nm in GRIN fibres keeping the same coupling conditions.}
	  \label{change_wavelength}
\end{figure}

We confirmed that spiral beams can always be obtained, also when the pump wavelength is varied from 1030 nm. In Fig.~\ref{change_wavelength}a and b we show the far-field at the output of GRIN and SI fibres, respectively, when the input wavelength ranges from 650 nm to 940 nm, while keeping the same incidence angle and position of the input beam. In Fig.~\ref{change_wavelength}c and d we report the corresponding near-field distributions and spectra for the case of a GRIN fibre only.
Again, we can remark that the formation of far-field spiral shapes is always observed, for any value of the source wavelength.
In Fig.\ref{change_wavelength}e we report the spectra obtained from GRIN MMFs, when varying the input peak power up to 49 MW. 

\bigskip
\subsection{Spiral emission from singlemode fibre and from laser pointer}
\begin{figure}
	  \centering
	  \includegraphics[width=0.75\textwidth]{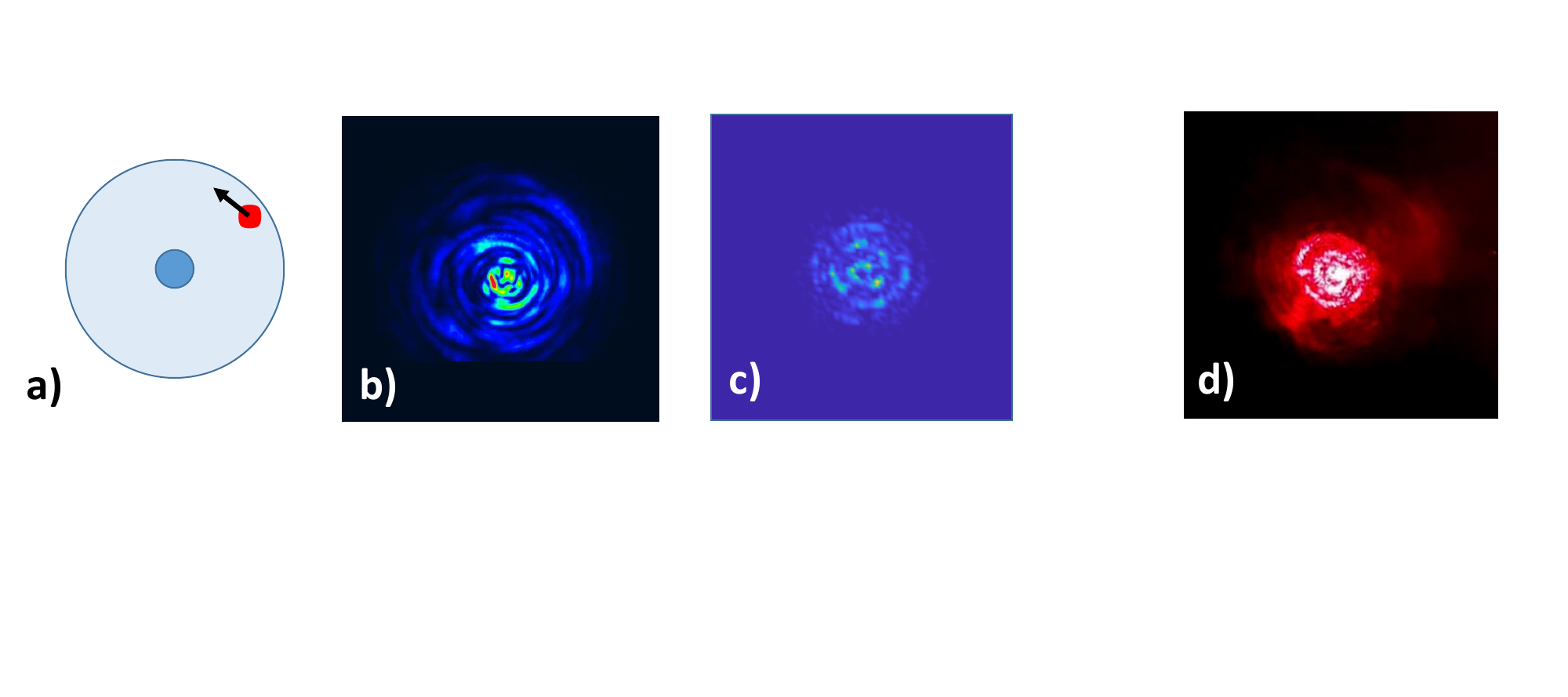}
	  \caption{(a) Sketch of the input coupling configuration for spiral emissions in an SMF. Black arrows represent the in-plane component of the input wave vector. (b) Experimental and (c) numerical far-field images from a 2 cm long SMF at 1030 nm. The input peak power is 0.1 MW with a repetition rate of 50 kHz, and the coupling angles are $\vartheta=2^{\circ}$ and $\varphi=45^{\circ}$. (d) Spiral emission from a commercial laser pointer.}
	  \label{spiralSMF}
\end{figure}

The objective of this section is to demonstrate, both experimentally and numerically, that spiral emission from a singlemode fiber (SMF) only occurs when injecting the beam inside the cladding. Fig.~\ref{spiralSMF}a schematically illustrates the input coupling condition. The laser beam is offset by about 45 $\mu$m with respect to the fibre axis, with $\vartheta=1.5^{\circ}$ and $\varphi=45^{\circ}$. Figs.~\ref{spiralSMF}b,c show the far-field image at the output of a 2 cm long SMF, as obtained either experimentally or numerically, under the same conditions. The wavelength and the pulse duration were kept at 1030 nm and 180 fs, respectively, while the input peak power was 0.1 MW, and the repetition rate was 50 kHz.
As can be seen from these images, even in the case of a singlemode core it is possible to generate spiral emission out of the fibre, by suitably coupling the incoming laser beam with the fibre cladding. 
Additionally, in Fig.~\ref{spiralSMF}d we report the far-field image from GRIN MMFs, using a commercial laser pointer.

